# Optimizing Coordinative Schedules for Tanker Terminals: An Intelligent Large Spatial-Temporal Data-Driven Approach - Part 2

*Deqing Zhai[a,b,\*], Xiuju Fu[a,\*], Xiao Feng Yin[a], Haiyan Xu[a], Wanbing Zhang[a] and Ning Li[a]*

[a] *1 Fusionopolis Way, Institute of High Performance Computing, Agency for Science, Technology and Research, Singapore 138632*

[b] *50 Nanyang Avenue, School of Electrical and Electronic Engineering, Nanyang Technological University, Singapore 639798*

## Abstract

In this study, a novel coordinative scheduling optimization approach is proposed to enhance port efficiency by reducing weighted average turnaround time. The proposed approach is developed as a heuristic algorithm applied and investigated through different observation windows with weekly rolling horizon paradigm method. The experimental results show that the proposed approach is effective and promising on mitigating the turnaround time of vessels. The results demonstrate that largest potential savings of turnaround time (weighted average) are around 17 hours (28%) reduction on baseline of 1-week observation, 45 hours (37%) reduction on baseline of 2-week observation and 70 hours (40%) reduction on baseline of 3-week observation. Even though the experimental results are based on historical datasets, the results potentially present significant benefits if real-time applications were applied under a quadratic computational complexity.

[\*]Corresponding author(s):

Email address (es): dzhai001@e.ntu.edu.sg (Deqing Zhai), fuxj@ihpc.a-star.edu.sg (Xiuju Fu)

# Introduction

Maritime logistics has been as a major and significant element in global supply chain networks for a quite long time. Due to its particular and big inertia in its domain, however, changes that mean to enhance its corresponding industry are quite difficult to make be adaptable. Therefore, little improvement over the industry would take a big leap on overall maritime supply chain networks globally. It is important to note that big potential enhancements could be made if cutting-edge machine-learning and optimization techniques were adopted. New solutions to old problems may inject new ways of thinking, and sometimes they also may turn our conventional mindsets into totally different ones with wider views.

As per International Maritime Organization (IMO) requirement in 2000, vessels ought to carry automatic identification system (AIS) devices on board. The device is able to broadcast static, dynamic and voyage/cargo related information to other vessels, shore authorities or relevant satellites. With the help of these devices, a big spatial-temporal data would be generated every few seconds with more than 50,000 vessels under operation all over the world at any moment. Given this big data, some hidden knowledge could be analyzed out, such as traffic risk analyses, route pattern analyses, abnormal analyses, etc. In this work, the scope is to identify conventional schedules based on historical AIS datasets, and further to propose a systematical heuristic approach to optimize the coordinative schedules for tankers and terminals with weekly rolling horizon paradigm method.

The paper is organized as follows: Section 2 reviews relevant literature regarding on scheduling optimization in maritime traffic. Section 3 formulates the problems that are to be tackled. The proposed methodology is followed up in Section 4. Section 5 presents the results and corresponding discussion on the basis of proposed methodology. Final conclusion, limitation and future direction are drawn in Section 6.

# Literature Review

In the study of Wang back in 1999, allocations of different cargo types from different mixture vessels in all directions are very challenging problems [1]. Current allocations are mostly still based on "First-Come, First-Served", which still have a lot of potential improvement possibilities. In order to efficiently resolve these problems, coordinative scheduling is one of necessary key processes. Different from long-term planning, coordinative scheduling focuses more on short-term scale, such as hours, days or weeks. This paradigm would not only locate feasible optimal schedules, but also effectively save computational efforts. As Nishimura et al. proposed in the work [2], which was using heuristic algorithm (genetic algorithm) to generate, simulate and evaluate the solutions. The feasible and optimal schedules were presented afterwards. However, the work presented only no more than 10 berths scenario. The problem sizes were quite less than those of real applications. As the computational efforts would exponentially increase with the growth of problem sizes, optimal solutions may not be reachable timely even with heavy computational efforts put when the problem sizes expand. Similarly, Zhang et al. proposed an approach that used simulated annealing with genetic algorithm coupled for 20 vessels. Again these simulated and optimal solutions were not able to fully represent the real application scenarios, and the computational efforts were relatively much larger in the proposed algorithm than its baseline approach [3]. For both work of Nishimura et al. and Zhang et al., their objective was to minimize the total wait time of vessels. Besides time aspect, there were other economic objectives under study as well, such as minimizing cost in study of [4] and maximizing profit in study of [5]. The scope of this study aims on minimizing time perspectives as direct focused objective.

To the best of our knowledge, most of scheduling optimizations in maritime domain are still based on container terminals and vessels in ports. Very few studies are about tanker terminals and vessels, probably due to their special cargo types and forms, complex operations, mobility constraints, non-standard volumes, etc. These difficult factors not only lead to hard problem formulations, but also tough optimizations. To address this type of hard problems, we propose our approach to tackle this type of problem on the basis of performance metric. Adopted from the study of Chung [6], this work uses average turnaround time as performance metric for overall vessels within observed windows.

## Problem Statement

Tanker scheduling optimizations are challenging problems due to their special natures of characteristics compared with container terminals and vessels. Based on the observations from historical data in 2017, numerous vessels and terminals were inefficient in wasting time on waiting for availability due to non-coordinative scheduling between each other. In this study, we propose a coordinative scheduling optimization by making certain preliminary assumptions based on real scenarios as follows: 1) certain tanker terminals are dedicated to certain vessels when visiting; 2) certain tanker terminals are inter-exchangeable when vessels visit; 3) certain tanker vessels are dedicated to certain fore-planned schedules; 4) certain tanker vessels are flexible to adapt available real-time schedules. By compromising all preliminary assumptions, coordinative scheduling optimizations can be realized and evaluated in this study.

## Methodology

In this section, methodology is illustrated in details, such as preliminary data statistics, pre-processing and cleaning, proposed approach and paradigm method in this study.

### Data Statistics

Before going deep into the proposed methodology, data statistics are overviewed in this section. Given AIS data and terminal information, one month data in May 2017 is selected and prepared for this study. The overview statistics are tabulated in Table 1 and the locations of anchorages and berths are presented in Figure 1.

Table 1: Data Statistics (May 2017)

| Item | Value | Remark |
| --- | --- | --- |
| Anchorage | 62 | Number of anchorages |
| Berth | 288 | Number of berths |
| Vessel | 1218 | Number of distinctive vessels |
| Portcall | 1628 | Number of distinctive portcalls |
| Datetime | 01 May 2017 | Starting at 00:00:00 |
| Datetime | 31 May 2017 | Ending at 23:42:00 |

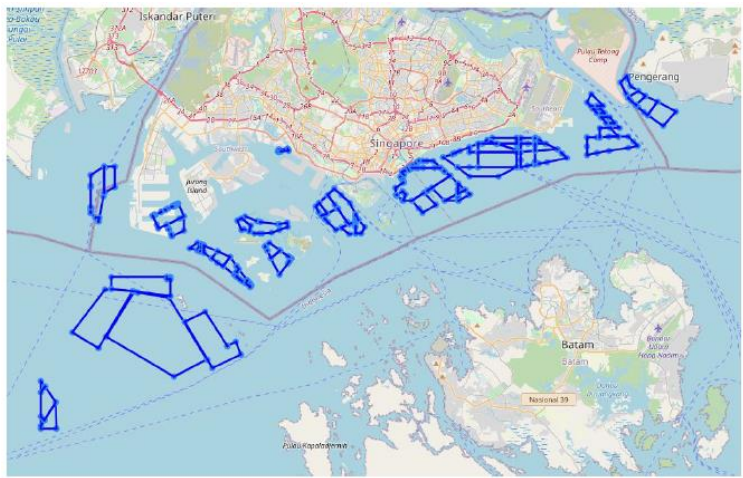

(a) Anchorage Areas

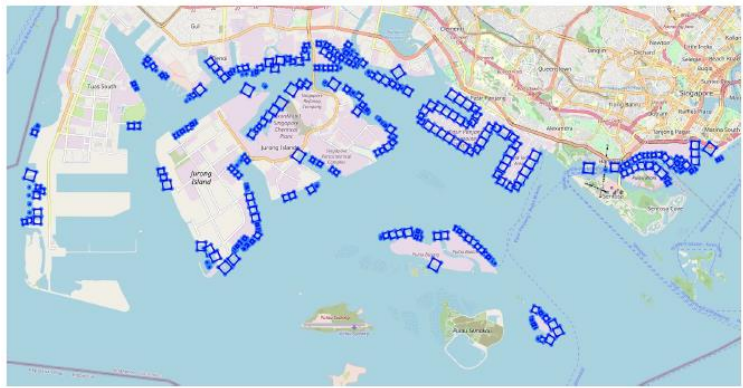

(b) Berth Areas

Figure 1: Locations of Anchorages and Berths

## Data Pre-processing

In order to have solid, substantial and comprehensive conclusions, raw data has to be pre-processed before plugging into optimization algorithms. The raw data is stored in a way of large amount of time series with spatial information. The spatial data is geographical longitude and latitude with respect to a designated timestamp. Since the raw data was originated from AIS devices equipped on board, errors are inevitably introduced by apparatus and human being, such as wrong signal recorded, AIS device purposely turned off by on-board crew, wrong information inputted by on-board crew, etc. Hereby, data pre-processing for standardizing and cleaning the raw data is one of necessary and essential works to complete first.

### Data Standardizing

In this study, the data extracted from database is standardized by information as follows:

a) Location of stay: This covers the latitude and longitude where vessels stay, such as anchorage or berth locations of waiting or conducting operations, respectively, e.g. (1.277109, 103.911774) or (1.276425, 103.663339).
b) Vessel identity: This includes maritime mobile service identity (MMSI) of vessels for differentiating each particular vessel, e.g. 220209000.
c) Timestamp of stay: This records the starting and ending timestamps of vessels staying at different specific locations, e.g. 2017-05-21 18:10:00 and 2017-05-22 03:04:00.
d) Fixedness of scheduled timing: This indicator reflects if the scheduled timing between vessels and terminals are fixed or flexible, e.g. 0 (flexible) and 1 (fixed).
e) Fixedness of scheduled location: This factor indicates if the scheduled locations between vessels and terminals are fixed or flexible, e.g. 0 (flexible) and 1 (fixed).

### Data Cleaning

In this study, the data cleaning is illustrated for some key and frequent issues listed as follows:

a) Location drifting: By analyzing the given AIS data, some AIS points are presented as far from their reasonable locations based on their near-term AIS data. This is called drifting in the context of this study as shown in Figure 2. These sudden changes in locations could trigger other errors that propagate in berth stay duration for follow-up processes. Therefore, intelligent algorithms are developed to classify if drifting occurs and output the cleaned data thereafter. Since this part is not the key scope of this work, the details of algorithms for drifting cleaning will be presented in follow-up future studies.

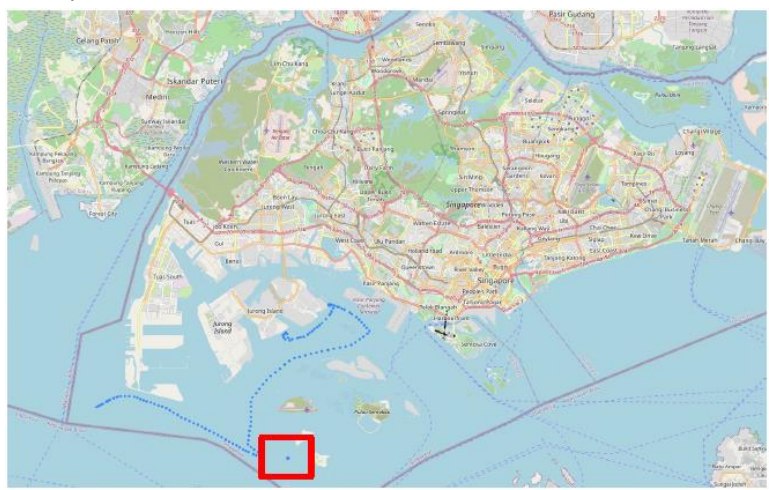

Figure 2: Example Case - Location Drifting

b) Location missing: In particular portcalls of vessels, on-board AIS devices are not always turned on by crews besides they malfunction, so some missing gaps could be observed as shown in Figure 3. Based on the missing parts and near-term AIS data, certain missing gaps could be confidently recovered from recovery algorithms, while others have to leave them as they are due to extremely unpredictable behaviors. Again this part is not the key scope of this work, the details of algorithms for missing gap recovery will be covered in future studies.

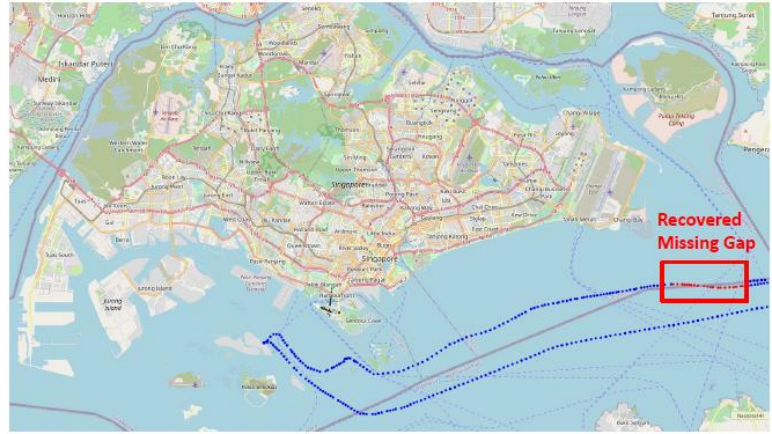

Figure 3: Example Case - Location Missing

## Optimization for Coordinative Scheduling

The main scope of this study is to present and evaluate the proposed approach on coordinative scheduling optimization. The coordinative factors are simulated by fixedness of scheduled timing and location. The fixedness is also incorporated with randomness with hyper-parameters $(T, S)$ introduced to simulate real application scenarios and examine how the performance of scenarios would be.

### Approach

Given the historical one month AIS data and knowledge of locations in port, historical activities and schedules are gathered. Based on these informative datasets, a two-step heuristic algorithm is proposed for coordinative scheduling optimization as shown in Figure 4.

```
Input: Initial Schedules, X
       Buffer timing, τ
Output: Optimal Schedules, X*

Step 1:
V_list ← getVList(X)
for each V in V_list:
    S_list ← getSchedules(V)
    for each S_i in S_list:
        if (flag_temporal_cur != 1):
            S_{i,t} ← S_{i-1,t} + τ
        if (flag_spatial_cur != 1):
            S_{i,s} ← S_{i-1,s}
        else:
            Remain unchanged.
        X^temp ← S_i

Step 2:
TB_list ← getTBList(X^temp)
for each TB in TB_list:
    S_list ← getSchedules(TB)
    for each S_i in S_list and S_list overlaps:
        if (flag_temporal_cur != 1):
            S_{i,t} ← S_{i-1,t} + τ
        else:
            if (flag_spatial_pre != 1):
                S_{i-1,s} ← Available(S_{list,s})
            else:
                if (flag_spatial_cur != 1):
                    S_{i,s} ← Available(S_{list,s})
                else:
                    Remain unchanged.
        X* ← S_i
```

Figure 4: Proposed Heuristic Algorithm (Pseudo-code)

During scheduling, a reasonable buffer time (τ) is preset 1 hour to link adjacent activities. The optimization is covered by different scenarios that are tuned by hyper-parameterized (T, S). Besides the heuristic algorithm, it is also important to note that data visibility or so called data observation window is also one of most significant factors in coordinative scheduling optimization. Based on our previous studies, it is necessary to have sufficient data visibility on future perspective to obtain optimal solutions to the scheduling problems. But there is an upper limit as well saying that the optimal solutions would have insignificant enhancement further beyond upper limit. In addition, the higher data visibility also requires the heavier computational efforts. Therefore, there always exists a trade-off point for real-time applications between optimity and efficiency of solution gathering. Thus, three different levels of data visibility are considered and evaluated in this work as well, such as 1-week, 2-week and 3-week observations, in order to draw comprehensive and solid conclusions.

### Scenario

With the introduced hyper-parameters (T, S), different types of scenarios could be achieved by manipulating this pair of hyper-parameters. Many scenarios could be investigated, but only some typical scenarios are incorporated with sweeping each hyper-parameter from 0.1 to 0.9. Some typical intuitive interpretations on these hyper-parameterized scenarios are listed as follows:

a) $(T, S) = (0.1, 0.1)$: This scenario interprets the real-time cases that are very rigid to the dedicated schedules on timing and location.
b) $(T, S) = (0.1, 0.9)$: This scenario interprets the real-time cases that are very rigid to the dedicated schedules on timing, but location has flexibility to be coordinately scheduled.
c) $(T, S) = (0.9, 0.1)$: This scenario interprets the real-time cases that are very rigid to the dedicated schedules on location, but timing has flexibility to be coordinately scheduled.
d) $(T, S) = (0.9, 0.9)$: This scenario interprets the real-time cases that are very flexible to be coordinately scheduled on timing and location.

## Result and Discussion

In this section, results and discussion on different scenarios are presented. Two aspects are studied in this work: (1) optimization effectiveness (i.e. the potential of reduction on weighted average turnaround time); (2) optimization time complexity. The first part evaluates whether the proposed approach has the capability of addressing the coordinative scheduling problem effectively, while the second part investigates whether this proposed approach has the advantages on near real-time performance.

### Optimization Effectiveness

As described in previous sections, the proposed heuristic algorithm is applied and incorporated with different hyper-parameterized scenarios and data visibility levels. Different data visibility levels are tabulated in Table 2 for one month data in May 2017 with respect to number of vessels in each weekly rolling paradigm.

Table 2: Number of Vessels for Weekly Rolling Paradigm Observation

| 1-Week Observation | | 2-Week Observation | | 3-Week Observation | |
|---|---|---|---|---|---|
| wk1 | 369 | wk1+wk2 | 504 | wk1+wk2+wk3 | 617 |
| wk2 | 350 | wk2+wk3 | 499 | wk2+wk3+wk4 | 596 |
| wk3 | 354 | wk3+wk4 | 493 | wk3+wk4+wk5 | 530 |
| wk4 | 355 | wk4+wk5 | 408 | - | - |
| wk5 | 193 | - | - | - | - |

As aforementioned, the experimental results of turnaround time saving (weighted average) are presented in Table 3 for 1-week observation of data visibility. The benchmark turnaround time (weighted average) for these five weekly rolling groups is 60.140 hours. The proposed approach performs proportionally better with higher hyper-parameterized scenarios with a saving result of about 17 hours (equivalent to 28% reduction on 1-week benchmark result) at the best.

Table 3: Turnaround Time Saving (weighted average) - Weekly Rolling by 1-Week Observation

| T/S | 0.1 | | 0.3 | | 0.5 | | 0.7 | | 0.9 | |
|-----|-------|-------|-------|-------|-------|-------|-------|-------|-------|-------|
|     | Hours | %     | Hours | %     | Hours | %     | Hours | %     | Hours | %     |
| 0.1 | 1.14  | 1.90  | 1.37  | 2.27  | 0.83  | 1.39  | 1.30  | 2.16  | 1.32  | 2.20  |
| 0.3 | 4.16  | 6.92  | 4.41  | 7.33  | 4.06  | 6.75  | 4.46  | 7.41  | 4.37  | 7.26  |
| 0.5 | 7.53  | 12.51 | 7.73  | 12.85 | 7.39  | 12.28 | 7.40  | 12.31 | 7.39  | 12.28 |
| 0.7 | 11.51 | 19.15 | 11.37 | 18.91 | 11.83 | 19.67 | 11.55 | 19.20 | 11.36 | 18.89 |
| 0.9 | 15.31 | 25.46 | 15.84 | 26.34 | 16.83 | 27.99 | 16.88 | 28.06 | 17.07 | 28.38 |

Note: T = Temporal flexibility; S = Spatial flexibility; for each vessel.
Benchmark turnaround time (weighted average) = 60.140 hrs

Similarly, the experimental results of turnaround time saving (weighted average) are presented in Table 4 for 2-week observation of data visibility. The benchmark turnaround time (weighted average) for these five weekly rolling groups is 120.993 hours. The proposed approach performs proportionally better with higher hyper-parameterized scenarios with a saving result of about 45 hours (equivalent to 37% reduction on 2-week benchmark result) at the best.

Table 4: Turnaround Time Saving (weighted average) - Weekly Rolling by 2-Week Observation

| T/S | 0.1 | | 0.3 | | 0.5 | | 0.7 | | 0.9 | |
|-----|-------|-------|-------|-------|-------|-------|-------|-------|-------|-------|
|     | Hours | %     | Hours | %     | Hours | %     | Hours | %     | Hours | %     |
| 0.1 | 1.93  | 1.60  | 2.82  | 2.33  | 2.74  | 2.26  | 2.06  | 1.70  | 1.79  | 1.48  |
| 0.3 | 8.22  | 6.79  | 8.81  | 7.28  | 8.92  | 7.38  | 8.71  | 7.20  | 8.81  | 7.28  |
| 0.5 | 17.27 | 14.27 | 16.72 | 13.82 | 17.63 | 14.57 | 17.45 | 14.42 | 16.31 | 13.48 |
| 0.7 | 27.97 | 23.12 | 27.93 | 23.09 | 29.27 | 24.19 | 28.13 | 23.25 | 29.49 | 24.38 |
| 0.9 | 45.41 | 37.53 | 44.56 | 36.83 | 42.83 | 35.40 | 45.03 | 37.22 | 44.53 | 36.81 |

Note: T = Temporal flexibility; S = Spatial flexibility; for each vessel.
Benchmark turnaround time (weighted average) = 120.993 hrs

Lastly, the experimental results of turnaround time saving (weighted average) are presented in Table 5 for 3-week observation of data visibility. The benchmark turnaround time (weighted average) for these five weekly rolling Groups is 180.223 hours. The proposed approach performs proportionally better with higher hyper-parameterized scenarios with a saving result of about 70 hours (equivalent to 40% reduction on 3-week benchmark result) at the best.

Table 5: Turnaround Time Saving (weighted average) - Weekly Rolling by 3-Week Observation

| T/S | 0.1 | | 0.3 | | 0.5 | | 0.7 | | 0.9 | |
| --- | --- | --- | --- | --- | --- | --- | --- | --- | --- | --- |
| | Hours | % | Hours | % | Hours | % | Hours | % | Hours | % |
| 0.1 | 2.90 | 1.61 | 3.95 | 2.19 | 4.80 | 2.66 | 3.27 | 1.81 | 3.02 | 1.68 |
| 0.3 | 11.89 | 6.60 | 14.53 | 8.06 | 12.48 | 6.92 | 13.01 | 7.22 | 12.57 | 6.98 |
| 0.5 | 25.07 | 13.91 | 24.99 | 13.87 | 25.43 | 14.11 | 25.95 | 14.40 | 25.80 | 14.31 |
| 0.7 | 41.38 | 22.96 | 43.96 | 24.39 | 43.08 | 23.91 | 43.31 | 24.03 | 43.62 | 24.20 |
| 0.9 | 69.29 | 38.45 | 71.92 | 39.91 | 70.86 | 39.32 | 71.64 | 39.75 | 72.80 | 40.40 |

Note: T = Temporal flexibility; S = Spatial flexibility; for each vessel.
Benchmark turnaround time (weighted average) = 180.223 hrs

With the experimental results shown above, it is important to note that temporal flexibility plays a dominant role in turnaround time reduction, while spatial flexibility seems have little impact on the reduction via the proposed approach in coordinative scheduling optimizations. A reasonable explanation of this phenomenon may be due to sufficient number of available terminals in the experiments, so the spatial term is not a key factor to influence the turnaround time of vessels. While temporal term does direct impacts on turnaround time of vessels by presetting the schedule timing as fixed in coordinative scheduling optimizations.

## Computational Complexity

Based on the previously well-defined approach and paradigm method, the computational complexity of step 1 in the proposed heuristic algorithm is quadratic, $O(n^2)$, with respect to the optimization problem sizes. Apparently, the computational complexity of step 2 is also quadratic, $O(n^2)$. Thus, the whole algorithm has a computational complexity of $O(n^2)$ at the worst cases.

## Conclusion

In this section, conclusion will be drawn, and limitations will be followed. Lastly, possible future directions will be pointed out.

### Research Conclusion

In this study, the goal is to minimize the weighted average turnaround time of historical datasets under a span of 1-month in May 2017. By addressing this problem, a novel intelligent heuristic approach is proposed. The proposed approach composes with heuristic algorithm and developed paradigm methods. For the paradigm method, weekly rolling horizon is validated and adapted from previous studies. The advancement of this paradigm is to further study the data visibility or so called observation window in this work, such as 1-week, 2-week and 3-week observations. Based on the experimental results over different hyper-parameterized scenarios, the proposed approach achieves remarkable optimal results under a quadratic level of computational complexity.

According to the investigations over different scenarios, the results demonstrate that largest potential savings of turnaround time (weighted average) are around 17 hours (~28%) reduction on baseline of 1-

week observation, 45 hours (~37%) reduction on baseline of 2-week observation and 70 hours (~40%) reduction on baseline of 3-week observation.

Large saving of turnaround time would directly benefit multiple parties or stakeholders in maritime logistics network, such as fuel saving, lead time saving for shipping company, chartering time saving for consignors and delivery time saving for consignees, etc. These benefits could eventually translate into profits, but the evaluation of benefits is out of scope in this study.

## Research Limitation

This work proposed a novel approach on optimizing coordinative scheduling problem and the approach is validated through different hyper-parameterized scenarios. However, limitations are still involved in this study.

a) Assumption: As defined in introducing hyper-parameters $(T, S)$, there is no distinction between terminals when schedule is optimized by transferring terminals to other available ones. However, this is an assumption for simplicity. In reality, certain terminals are particularly non-interchangeable.
b) Data: In maritime logistics network, different types of vessels are voyaging in Singapore Strait. However, the data used for this study is merely from tanker vessels. This will limit the overall picture of network in Singapore Strait. But current dataset still successfully demonstrates the capability of acquiring large potential benefits.

## Future Direction

Maritime logistics network will still play an important role in worldwide transport and economy. Global ports are developing towards a direction of high-efficiency and high-smartness. The next step of study would be further cultivating maritime domain, for instance, incorporating other types of vessels that have significant impacts on port efficiency and congestion. On the other side, optimization algorithms on solving scheduling problems could be further enhanced in terms of both effectiveness and computational efforts under different scenarios. The scenarios are also required to be fine-tuned further towards reality.

## Acknowledgment

The authors would like to thank Mr. Chua Chye Poh and many others from ShipsFocus Group (https://www.shipsfocus.com) for providing helpful discussion and domain knowledge in the field of maritime. The authors also gratefully thank for all kinds of reviews, constructive comments and suggestions.

# Appendix

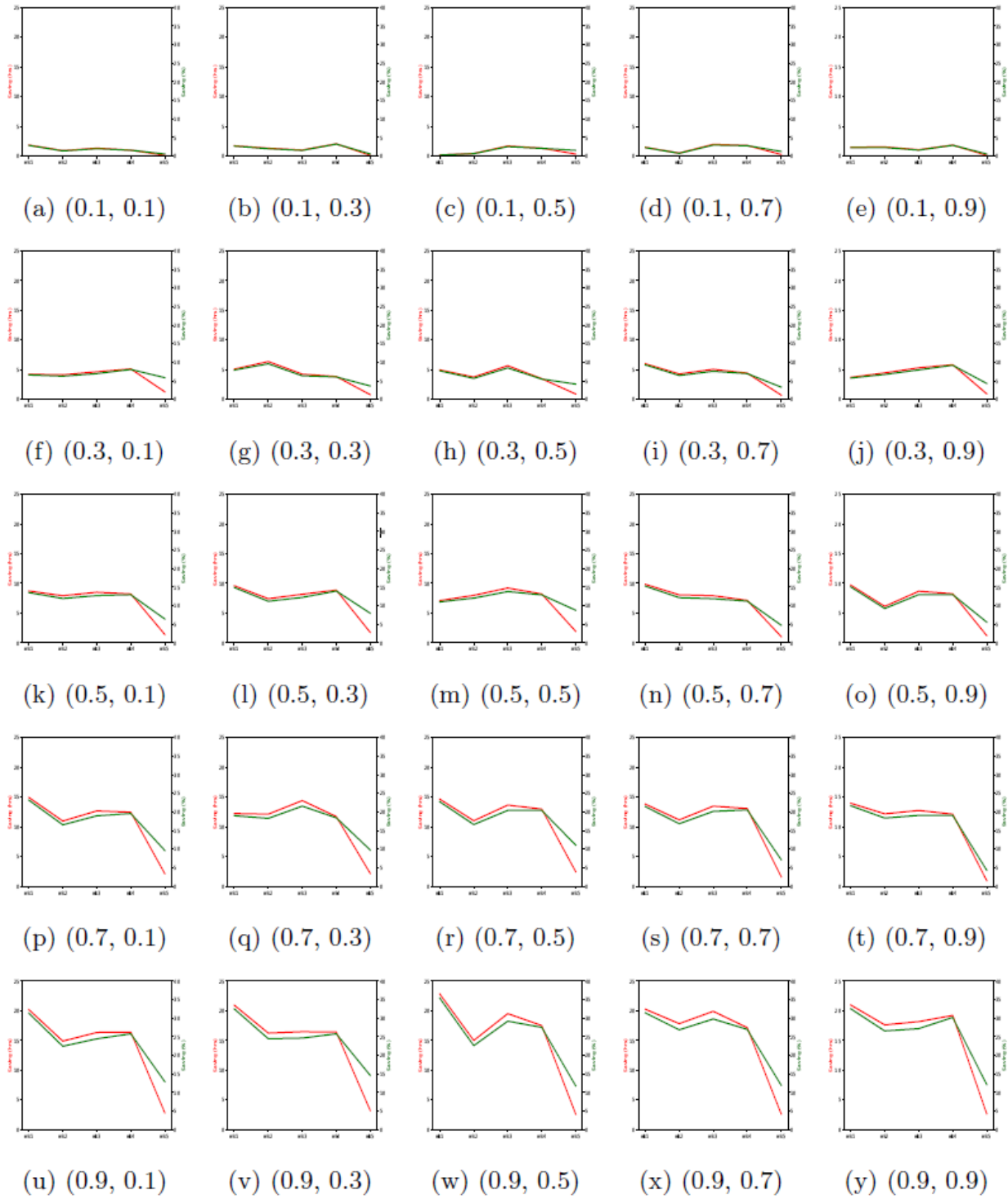

Figure 5: Results of Weekly Rolling by 1-Week Observation

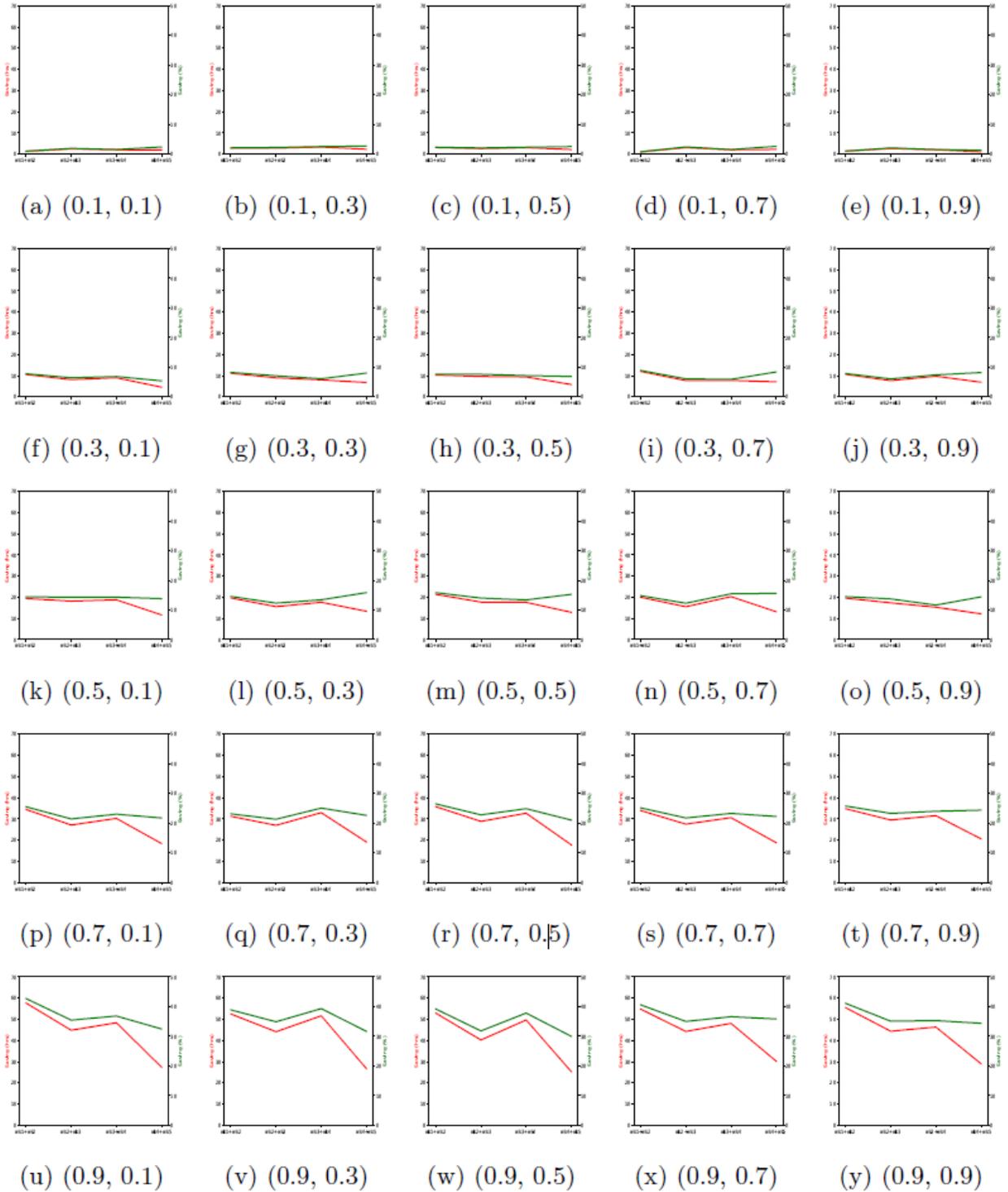

Figure 6: Results of Weekly Rolling by 2-Week Observation

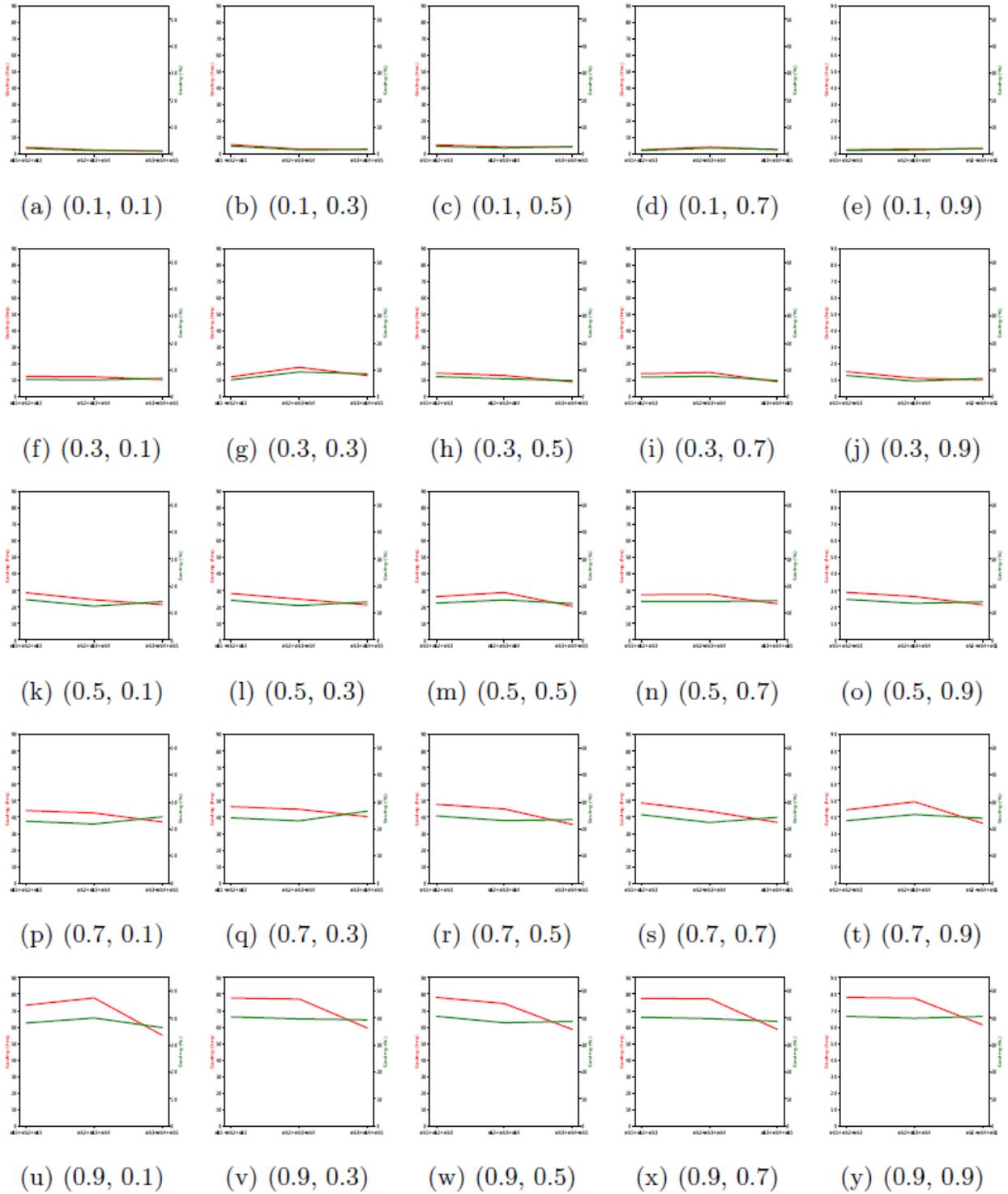

Figure 7: Results of Weekly Rolling by 3-Week Observation